# PDM BASED I-SOAS DATA WAREHOUSE DESIGN

**Zeeshan Ahmed**
Vienna University of Technology, Getreidemarkt 9/307 1060, Vienna, Austria.
Email: zeeshan.ahmed@tuwien.ac.at

## ABSTRACT

This research paper briefly describes the industrial contributions of Product Data Management in any organization's technical and managerial data management. Then focusing on some current major PDM based problems i.e. Static and Unintelligent Search, Platform Independent System and Successful PDM System Implementation, briefly presents a semantic based solution i.e. I-SOAS. Majorly this research paper is about to present and discuss the contributions of I-SOAS in any organization's technical and system data management.

## 1. INTRODUCTION

*Product Data Management* (PDM) is a digital electronic way of maintaining organizational data to maintain and improve the quality of products and followed processes. PDM based products mainly maintain the information about the organization including personal involved in managerial and technical operations, running projects and manufacturing products [1]. Where PDM based products are heavily benefiting industry there PDM community is also facing some serious unresolved issues i.e. successful secure platform independent PDM system implementation, PDM system deployment and reinstallation, static and unfriendly machine interface, unintelligent search and scalable standardized framework.

Many approaches and solutions systems including Meta-phase (SDRC), SherpaWorks (Inso), Enovia (IBM), CMS (WTC), Windchill (PTC), and Smarteam (Smart Solutions) [2] are proposed targeting these above mentioned problem oriented issues all together and on individual basis but still there is no as such one promising approach or product exists which claims of providing all the solutions.

Targeting some of above mentioned PDM based problem oriented issues i.e. Static and Unfriendly Graphical User Interface, Static and Unintelligent Search, Platform Independent System and Successful PDM System Implementation, we have also proposed an approach called Intelligent Semantic Oriented Search (I-SOAS) [3] (See Fig 1).

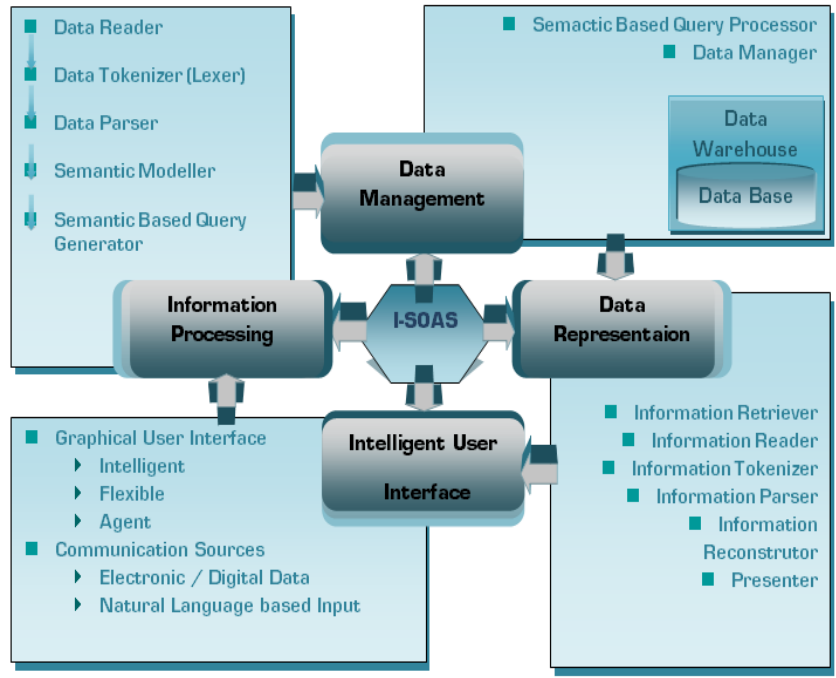

**Fig.1.** I-SOAS Conceptual Architecture [3]

I-SOAS is an agent, information engineering & modeling, data warehousing and knowledge base approach, proposed to provide solution in implementing a semantic based intelligent application capable of handling user's structured and unstructured requests by processing, modeling and managing the into data base. To meet aforementioned goals I-SOAS's proposed conceptual architecture is divided into four sequential iterative components i.e. Intelligent User Interface (IUI), Information Processing (IP), Data Management (DM) and Data Representation (DR) (See Fig 1). IUI is proposed to design intelligent human machine interface for system user communication, IP is proposed to process and model user's unstructured and structured inputted request by reading, lexing, parsing, and semantic modeling, DM is proposed to manage user request and system performance based information and DR is proposed to represent system outputted results in user's understandable format [4].

To implement I-SOAS we have designed an implementable architecture consisting of four main modules i.e. I-SOAS Graphical Interface, I-SOAS Data Warehouse, I-SOAS Knowledge Base, I-SOAS Processing Modelling and three communication layers i.e. Process Presentation Layer (PPL), Process Database Layer (PDL), and Process Knowledge Layer (PKL) (See Fig 2) [5].

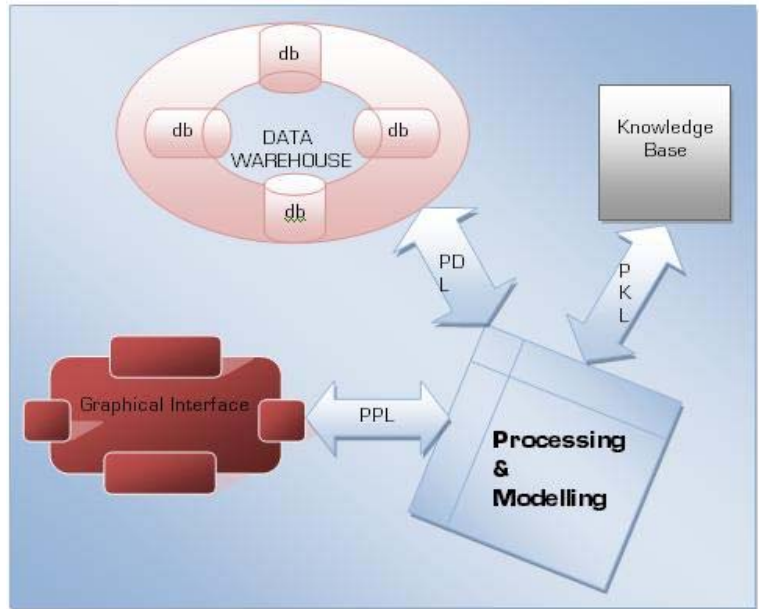

**Fig.2.** I-SOAS Implementable Architecture [5]

I-SOAS Graphical Interface is proposed to implement intelligent human machine interface, I-SOAS Data Warehouse is proposed to load and manage the organizational technical and managerial data, I-SOAS Knowledge Base is proposed to capture, manage, improve and deliver knowledge and ISOAS Processing and Modelling is proposed to read, organized, tokenize, parse, semantically evaluate and generate a semantic based queries to extract desired results from I-SOAS Data warehouse and I-SOAS Knowledge base. Three communication layers are proposed to transfer data between above mentioned four major modules of I-SOAS Implementable Architecture (See Fig 2).

In this research we are not going in details of any other component of I-SOAS except I-SOAS Data Warehouse. Initially we present I-SOAS Data Warehouse in detail in section 2 of this research paper, then we present I-SOAS Data Warehouse design requirements in section 3, I-SOAS Data Warehouse entity relationship design in section 4 and I-SOAS Data Warehouse's system sequence design in section 5. Moreover we also present information about involved tools and technologies in the development of I-SOAS Data Warehouse in section 6, conclude discussion in section 7 and present the future recommendation in section 8 of this research paper.

## 2. I-SOAS DATA WAREHOUSE

The main theme or the idea behind the I-SOAS Data Warehouse is to develop a repository to store and manage Product Data Management based Application's heavy volume data. Moreover the designed I-SOAS Data warehouse is conceptually based and meeting the requirements of third component of I-SOAS's conceptual iterative architecture Data Management (see Fig 1).

As we said before, I-SOAS Data Warehouse module is manly the repository to store, extract, transform, load and manage organizational technical and managerial data. I-SOAS Data Warehouse is supposed to work like other Data Warehouses by providing several options to produce common data model for all data of interest easier to report and analyze information, prior loading data, security and retrieval of data without slowing down operational systems [5].

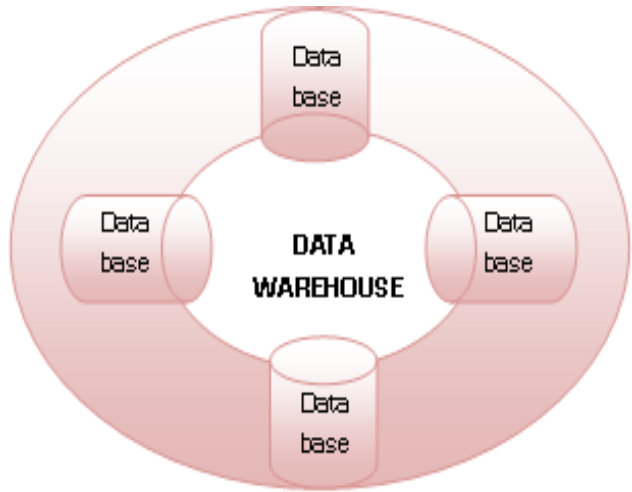

**Fig 3.** I-SOAS: Data Warehouse

The technical architecture of I-SOAS Data Warehouse consists of five main steps i.e. Extraction, Transformation, Loading, Security and Job control [6].
1. Data will be extracted from different sources during Extraction
2. Data will be transformed by management, integration, de-normalization, cleansing, conversion, aggregation, and auditing in transformation.
3. Data will be loaded in cyclic process in Loading.
4. Data will be secured in security.
5. Job scheduling, monitoring, logging, exception handling, error handling, and notification will be performed in Job handling.

### 3. I-SOAS DATA WARE DESIGN REQUIREMENTS

The I-SOAS Database must be designed keeping two major record keeping requirements i.e. *Organizational Data and System Data* in mind.

**A) Organizational Data**

The designed database must be capable of storing and managing organizational technical and non technical data divide into three sub major categories *User Data, Project Data and Product Data*.

The designed database must be capable of storing and managing user data containing user's personal information, role(s), responsibilities, type (s), group(s) and right(s) etc.

The designed database must be capable of storing and managing Project data containing detailed information about project, start end dates, actions, ownerships, category, state, tasks, meeting, status and deadlines etc.

The designed database must be capable of storing and managing Product data containing detailed information about the product produced in the organization, release dates and important notes etc.

**B) System Data**

The designed database must be capable of storing and managing System Data containing the information about user inputs, system output, actions, reactions, input processing, modeling and event log etc.

## 4. I-SOAS DATA WAREHOUSE ERD

Following the designed requirements two entity relationship diagram are designed for I-SOAS Database i.e. *I-SOAS Database Organizational Data and I-SOAS Database System Data*.

**A) I-SOAS Database Organizational Data Design**

The designed entity relationship diagram of I-SOAS Database Organizational Data is consists of 5 main relations i.e. *Organization, Person, Staff, Project, Product* and 17 supportive relations i.e. *Name, Contacts, Contacts_Web, Contact_Telephone, Contact_City, City_Country, City_State, Start_End_Date, Project_Team, Meeting, Activity, Staff_Meeting, Document, Type, Staff_Document, Organisation_Document, Project_Document* designed and connected to store and manage organizational data, employee's (user) personal data, project data and product data.

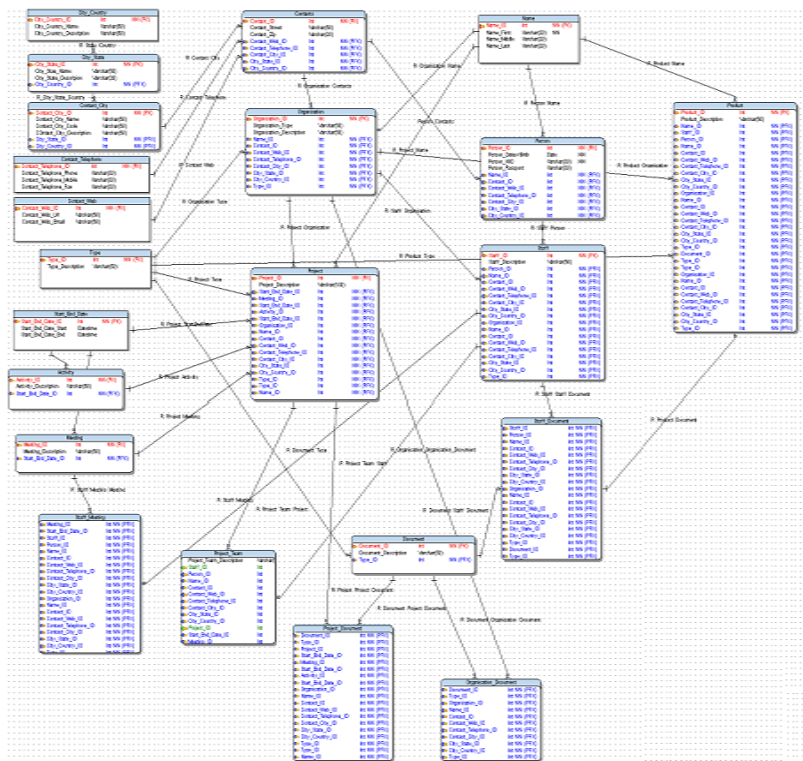

**Fig 4.** I-SOAS: Data Warehouse Organizational Design

- *Organisation* relation is designed to store and main the information about respective Organisation.
- *Person* relation is designed to store and main the information about every associated person with organization.
- *Staff* relation is designed to store and maintain the information of each and every employee of the organization
- *Project* relation is designed to store and maintain the information of each and every running is the organization.
- *Product* relation is designed to store and maintain the information of each and every product developed in the organization.
- *Document* relation is designed to store and maintain the information of each organizational personal, staff and project document.
- *Staff_Document* relation is an extended relation of relation Document and designed to store and maintain the joint information about staff and document.
- *Organisation_Document* relation is an extended relation of relation Document and designed to store and maintain the joint information about organization and document.
- *Project_Document* relation is an extended relation of relation Document and designed to store and maintain the joint information about project and document.
- *Name* relation is designed to store and maintain the information of the name of Organization, Person, Project and Product
- *Contacts* relation is designed to store and maintain the information of the contacts of Person and Organisation.
- *Contacts_Web* relation is an extended relation of relation Contact and designed to store and maintain the information about the web contacts consisting of email address and url.
- *Contact_Telephone* relation is an extended relation of relation Contact and designed to store and maintain the information about the telephone contacts consisting of Mobile, Fax and Telephone numbers.
- *Contact_City* relation is an extended relation of relation Contact and designed to store and maintain the information about the city contacts consisting of City Name and Code.
- *City_State* relation is an extended relation of relation City and designed to store and maintain the information about the belonging State contacts of the city.
- *City_Country* relation is an extended relation of relation City and designed to store and maintain the information about the belonging Country contacts of the city.
- *Project_Team* relation is designed to store and maintain the information of each team in the organization working on any organizational project.
- *Activity* relation is designed to store and maintain the information of each activity performed in the orgazation.
- *Meeting* relation is designed to store and maintain the information of each meeting between organizational staff.
- *Staff_Meeting* relation is an extended relation of relation Meeting and designed to store and maintain the joint information about staff (attended meeting) and meeting itself.
- *Type* relation is designed to store and maintain the information of the types of organization, document, project, product, system input and system output.

- *Start_End_Date* relation is designed to store and maintain the information of the start and end date and timing of organizational projects and meetings.

### B)  I-SOAS Database System Data Design

The designed entity relationship diagram of I-SOAS Database Organizational Data is consists of three relations i.e. *Login, SystemInput and SystemOutput*.

- *Login* relations is designed to store and maintain the user system login information (User name and Password).
- *SystemInput* relation is designed to maintain user input instruction by storing information about user inputted instruction details including time, date and user (who inputted).
- *SystemOutput* relation is designed to maintain system Output by storing information about system outputted instruction details including time, date, and respective input.

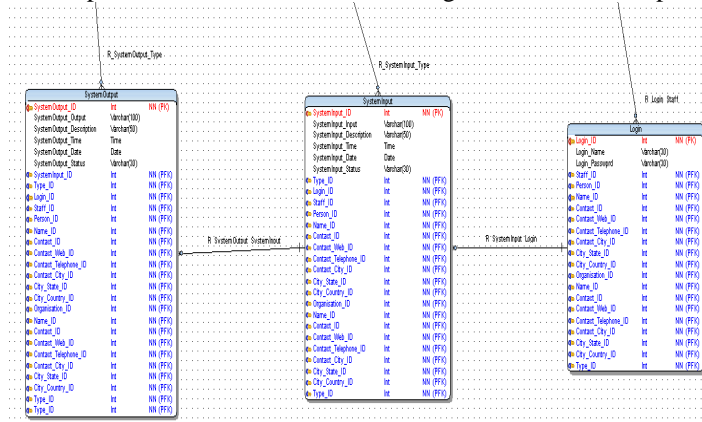

**Fig 5.** I-SOAS: Data Warehouse System Data

### 5. I-SOAS DATA WAREHOUSE SYSTEM SEQUENCE DESIGN

I-SOAS Data Warehouse System Sequence Design is consists of three main components i.e. Process Database Layer (PDL), Data Input and Database.

These three components are supposed to perform certain jobs. The job of Process Database Layer (PDL) is to bring system data from I-SOAS Processing and Modelling and forward to Data Input.  Then system data will be stored and managed in database by Data Input component. Then the final acknowledgement will be send to I-SOAS Processing and Modelling via Process Database Layer (PDL).

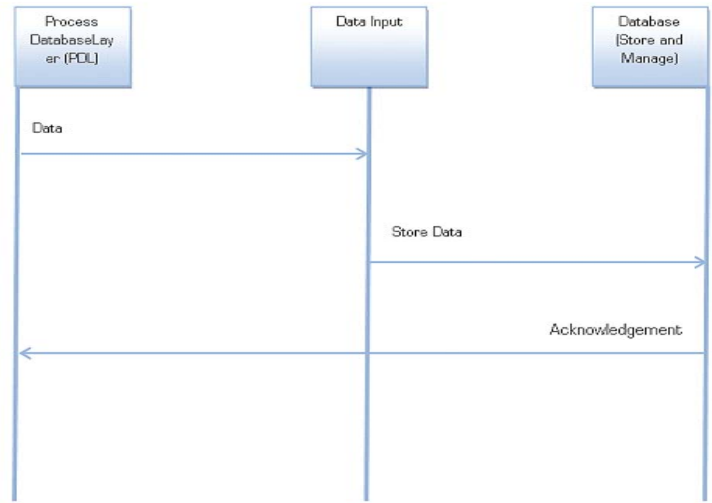

**Fig 6.** I-SOAS: Data Warehouse System Sequence Design

## 6. I-SOAS DATA WAREHOUSE TECHNOLOGIES INVOLVED

For the implementation of I-SOAS Data Warehouse we are considering MySQL Data Warehousing platform.

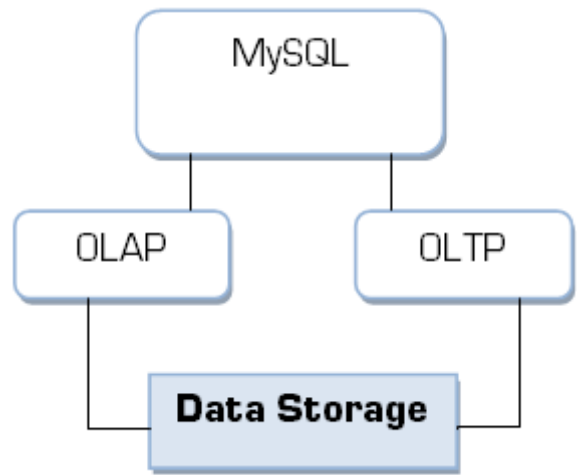

**Fig 7.** I-SOAS: Data Warehouse Technologies Involved

## 7. CONCLUSION

In this research paper we have briefly described Product Data Management and its some major existing challenges, then continuing the presentation of research briefly described the conceptual and implementable architecture of our own proposed solution based approach I-SOAS targeting PDM challenges. In this research paper we have also described the design requirements and implementation of I-SOAS Data Warehouse (a component of I-SOAS). Moreover describing the detailed information about I-SOAS Data Warehouse, we have presented information about the theme, design requirements, designed designs and technologies involved in the development of I-SOAS Data Warehouse.